\documentclass[useAMS,usenatbib]{mn2e}

\usepackage{graphicx}

\title[Nucleosynthesis of $^{56}$Ni in wind-driven Supernova Explosions]
{Nucleosynthesis of $^{56}$Ni in wind-driven Supernova Explosions and 
Constraints on the Central Engine of Gamma-Ray Bursts}
\author[K. Maeda and N. Tominaga]{Keiichi Maeda$^{1}$\thanks{E-mail:
keiichi.maeda@ipmu.jp} and 
Nozomu Tominaga$^{2}$\\
$^{1}$Institute for the Physics and Mathematics of the Universe (IPMU), University of Tokyo, 
Kashiwano-ha 5-1-5,\\ 
Kashiwa-shi, Chiba 277-8568, Japan\\
$^{2}$Division of Optical and Infrared Astronomy, National Astronomical Observatory of Japan, 
2-21-1 Osawa,\\ 
Mitaka, Tokyo 181-8588, Japan}
\begin{document}

\date{}

\pagerange{\pageref{firstpage}--\pageref{lastpage}} \pubyear{2008}

\maketitle

\label{firstpage}

\begin{abstract}
Theoretically expected natures of a supernova driven by a wind/jet 
are discussed. 
Approximate analytical formulations are derived to clarify basic 
physical processes involved in the wind/jet-driven explosions, and
it is shown that the explosion properties are characterized by the energy
injection rate ($\dot E_{\rm iso}$) and the mass injection rate ($\dot M_{\rm iso}$). 
To explain observations of SN 1998bw associated with Gamma-Ray Burst (GRB) 980425, 
the following conditions are required: 
$\dot E_{\rm iso} \dot M_{\rm iso} \ga 10^{51}$ erg $M_{\odot}$ s$^{-2}$ 
and $\dot E_{\rm iso} \ga 2 \times 10^{52}$ erg s$^{-1}$ 
(if the wind Lorentz factor $\Gamma_{\rm w} \sim 1$) or 
$\dot E_{\rm iso} \ga 7 \times 10^{52}$ erg s$^{-1}$ (if $\Gamma_{\rm w} \gg 1$).  
In SN 1998bw, $^{56}$Ni ($\sim 0.4M_{\odot}$) is probably produced in the shocked 
stellar mantle, not in the wind.  
The expected natures of SNe, e.g., ejected $^{56}$Ni masses and
ejecta masses, vary depending on $\dot E_{\rm iso}$ 
and $\dot M_{\rm iso}$. The sequence of the 
SN properties from high $\dot E_{\rm iso}$ and $\dot M_{\rm iso}$ to low $\dot E_{\rm iso}$ 
and $\dot M_{\rm iso}$ is the following: SN 1998bw-like -- intermediate case 
-- low mass ejecta ($\la 1M_{\odot}$) where $^{56}$Ni is from the wind -- whole collapse.  
This diversity may explain the diversity of supernovae associated with GRBs. 
Our result can be used to constrain natures of the wind/jet, 
which is linked to the central engine of GRBs, by studying properties of 
the associated supernovae. 
\end{abstract}


\section{Introduction}

Gamma-Ray Bursts (GRBs) are energetic cosmological events, emitting
$\ga10^{51}$ ergs in $\gamma$-ray. A leading model for the central engine 
of GRBs is the formation of a black hole (BH) and an accretion disk, 
following the gravitational collapse of a massive star whose main sequence mass 
($M_{\rm ms}$) is at least as large as $25M_{\odot}$ (for reviews, see 
Woosley \& Bloom 2006; Nomoto et al. 2007). 
A relativistic flow generated by neutrino annihilation (Woosley 1993; MacFadyen \& Woosley 1999)
or magnetic activity (Brown et al. 2000; Proga et al. 2003) 
is proposed to trigger a GRB. 

A link between (a class of) GRBs and Type Ic supernovae 
(SNe Ic) has been established observationally. 
The most convincing cases for the supernovae associated with 
GRBs (hereafter GRB-SNe) have been provided by spectroscopic detection of 
supernova features in an optical afterglow of a GRB or at the position 
consistent with a GRB. 
Three nearby GRB-SNe detected in this way are found to be similar to one another. 
The category includes GRB 980425/SN 1998bw (the proto-typical GRB-SN; 
Galama et al. 1998), 
GRB 030329/SN 2003dh (Hjorth et al. 2003; Kawabata et al. 2003; 
Matheson et al. 2003; Stanek et al. 2003), 
and GRB 031203/SN 2003lw (Malesani et al. 2004; Thomsen et al. 2004). 
Optical observations of these GRB-SNe are well explained by an explosion of a 
carbon-oxygen (CO) star, 
which has evolved from a massive star ($M_{\rm ms} \sim 40M_{\odot}$) and 
has lost its H- and He-envelopes during the hydrostatic evolutionary phase 
(Iwamoto et al. 1998; Woosley et al. 1999; Nakamura et al. 2001a; 
Mazzali et al. 2003, 2006). 
The kinetic energy ($E_{\rm K}$) of the expansion is large, 
$E_{51} \equiv E_{K}/10^{51}$ ergs $\ga 10$ (note that $E_{51} \sim 1$ 
for canonical supernovae). 
They eject $\sim 0.3 - 0.7M_{\odot}$ of $^{56}$Ni 
(which powers the SN luminosity by the decay chain $^{56}$Ni $\to$ Co $\to$ Fe). 
Hereafter, the mass of $^{56}$Ni is denoted by $M$($^{56}$Ni). 
Recently, another example of the association has been reported (Della Valle et al. 2008; 
Soderberg et al. 2008), i.e., GRB 081007/SN Ic 2008hw, while 
the observed properties of this SN have not been modeled yet. 

Despite the similarity within the well studied cases mentioned above, 
GRB-SNe do seem to have diverse properties. 
Peak magnitudes of so-called supernova bumps seen in GRB optical afterglows show 
diversity (Zeh, Klose, \& Hartmann 2004; Woosley \& Bloom 2006), 
highlighted by sub-luminous (possible) SNe in 
GRBs 040924 and 041006 (Soderberg et al. 2006). 
A few GRBs show no evidence for the supernova bump (Hjorth et al. 2000; Price et al. 2003). 
Non-detection of SN features in two nearby GRBs 060505 and 
060614 has been reported, 
placing the upper limit to brightness of possible underlying 
SNe $\sim 100$ times fainter than SN 1998bw 
(Della Valle et al. 2006; Fynbo et al. 2006; Gal-Yam et al. 2006). 

In spite of the observational constraints, the explosion mechanism of GRBs and GRB-SNe are still unknown.
In particular, how properties of the central engine are related to 
the bulk expansion of the stellar materials, observed as a supernova 
mainly in visual light, 
is still under debate. 
A possibility is that a supernova is induced by a disk wind 
generated by viscous heating (MacFadyen \& Woosley 1999; 
Narayan, Piran, \& Kumar 2001). 

Although there were numerical calculations for 
the wind/jet-driven explosions (Khokhlov et al. 1999; 
MacFadyen, Woosley, \& Heger 2001; 
Maeda \& Nomoto 2003; Nagataki et al. 2003; Maeda 2004, 
Nagataki, Mizuta, \& Sato 2006; Tominaga 2007a; Tominaga et al. 2007b; 
and Tominaga 2009), it has not been clarified what fundamentally determines 
the properties of resulting SNe and in what ways. 
Also, the numerical investigations have been restricted 
in the parameter space. 

Aiming to overcome these problems, this study is complementary to the 
past numerical studies. 
Our goal in this paper is to express theoretically expected 
features of SNe resulting from the wind/jet-driven explosion, 
as a function of rates of the mass and energy 
($\dot M_{\rm w}$ and $\dot E_{\rm w}$, where 
the subscript "w" denotes "wind", or $\dot M_{\rm iso}$ and 
$\dot E_{\rm iso}$, referring to the isotropic equivalent values) generated and 
injected from the central system (i.e., a black hole plus a disk) 
into the surrounding stellar mantle 
(in this paper, stellar "mantle" refers to the stellar materials above 
the central remnant, i.e., the outermost layer of the Fe core, 
and the Si- and CO-layers). 

Our strategy is the following. 
(1) We first clarify what are main ingredients of 
the wind/jet-driven SN explosion (\S 2). 
We develop a simplified description for the shock propagation and 
nucleosynthesis in the explosion (\S 3). 
We especially focus on the production of $^{56}$Ni, 
addressing how two proposed sites for the $^{56}$Ni production, 
a shocked stellar mantle and a disk wind (MacFadyen \& Woosley 1999; 
see Maeda \& Nomoto 2003 for a review), can be distinguished.
(2) We then compare the results with observations of SNe 
associated with a GRB, in order to 
constrain $\dot M_{\rm iso}$ and $\dot E_{\rm iso}$ in these SNe (\S 4). 
The required values for $\dot M_{\rm iso}$ and $E_{\rm iso}$ then 
should be regarded as conditions that any models for the 
central engine should satisfy. In other word, 
by constraining $\dot M_{\rm iso}$ and $\dot E_{\rm iso}$, 
we aim to provide useful constraints in studying 
the properties of the central engine of GRBs. 

\section{Models}

\begin{figure*}
   \centering
   \includegraphics[width=0.6\textwidth]{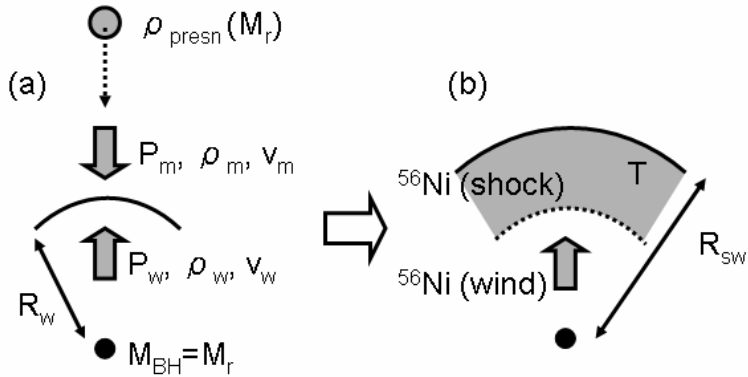}
   \caption{A schematic picture of a wind/jet-driven SN explosion. 
(a) A situation before the launch of the outward shock wave. 
The wind/jet is assumed to be continuously injected 
from the central remnant with properties denoted by the 
subscript "w" at radius $R_{\rm w}$. 
When materials initially at $M_{r}$ in the presupernova progenitor 
fall to $R_{\rm w}$, their properties are denoted by 
the subscript "m". At this point, the mass of the central remnant 
(the mass within $R_{\rm w}$) is $M_{r}$. 
(b) A situation after the launch of the outward shock wave. The shock wave 
arrives at radius $R_{\rm sw}$, and temperature behind the shock is 
denoted by $T$. $^{56}$Ni is produced both in the 
shocked stellar materials between $R_{\rm sw}$ and $R_{\rm w}$ and within the 
wind/jet itself.}
   \label{fig1}
\end{figure*}

We consider a situation that a supernova explosion 
is driven by outflow of materials (e.g., a disk wind; MacFadyen \& Woosley 1999) 
from the vicinity of a central remnant (likely a black hole). 
This energy input can be different from the relativistic jet 
producing a GRB; The wind can either be relativistic or non-relativistic 
at its injection from the central remnant into the surrounding stellar mantle. 

A schematic picture of the problem considered in this paper 
(and most of the past numerical studies) is shown in Figure 1. 
Throughout this paper, 
we adopt $25M_\odot$ and $40M_\odot$ progenitor models from Nomoto 
\& Hashimoto (1988). Following the treatment of the past numerical 
studies, the wind/jet (we
hereafter frequently call it simply a wind) is injected by hand at a
certain radius ($R_{\rm w}$ in this paper).

\subsection{Properties of the wind/jet}

In a one-dimensional hydrodynamic problem, 
three independent variables must be specified (two for thermodynamic variables 
and one for a hydrodynamic variable) as initial/boundary conditions.
For simplicity, we assume that the wind is dominated by the kinetic energy. 
This does not drastically defeat our results, since it is expected that 
the thermal energy deposited at the root of the wind, near the central remnant, 
is quickly converted to the kinetic energy well below $R_{\rm w}$. 
Thanks to this simplification, only two independent 
variables at $R_{\rm w}$ are required to determine the hydrodynamic evolution of the system. 
The choice of the independent variables can be arbitrary, thus we take 
the energy injection rate ($\dot E_{\rm w}$) and the mass injection rate 
($\dot M_{\rm w}$). 
The Lorentz factor of the wind/jet at the injection (at $R_{\rm w}$) 
is expressed as a function of these 
two independent variables, i.e., 
$\Gamma_{\rm w} \sim 1 + \dot E_{\rm w}/(\dot M_{\rm w} c^2)$, 
where $c$ is the speed of light. 
Hereafter, quantities expressing the properties of the wind 
at injection (at $R_{\rm w}$) are denoted by the subscript "w". 
Trivially, the wind is initially highly relativistic only if 
$\dot E_{{\rm w}, 51}/\dot M_{{\rm w}, \odot} \gg 1000$, 
where $\dot E_{{\rm w}, 51} \equiv \dot E_{\rm w}/(10^{51}$ erg s$^{-1})$ 
and $\dot M_{{\rm w}, \odot} \equiv \dot M_{\rm w}/(M_{\odot}$ s$^{-1})$. 

Our analysis is based on a one-dimensional radial flow; 
The collimation of the wind is taken into account with a
geometrical factor $f_{\Omega} (\le 1)$.
$f_{\Omega}$ relates the wind 
intrinsic properties and isotropic equivalents as 
\begin{eqnarray}
\dot E_{\rm w} & = & f_{\Omega} \dot E_{\rm iso} \ , \\
\dot M_{\rm w} & = & f_{\Omega} \dot M_{\rm iso} \ , \\
E_{\rm w} & = & f_{\Omega} E_{\rm iso} \ ,
\end{eqnarray} 
where $E_{\rm w}$ is the intrinsic total energy (i.e., the injected energy 
integrated over time), and the subscript "iso" refers to isotropic 
equivalent values. 
$f_{\Omega}$ is a measure of 
the collimation angle of the wind at $R_{\rm w}$: 
When the wind is a spherically symmetric flow, $f_{\Omega} = 1$. 
The wind is more narrowly collimated at the injection for smaller $f_{\Omega}$. 

For the temporal history of the wind properties, we focus on the situation that the 
wind is injected at $R_{\rm w}$, with $\dot E_{\rm w}$ and $\dot M_{\rm w}$ 
constant in time for $t < t_{\rm w}$, and then terminated at $t = t_{\rm w}$, so that 
\begin{equation}  
E_{\rm w} = \dot E_{\rm w} t_{\rm w} \ ({\rm or} \ {\rm equivalently,} \ 
E_{\rm iso} = \dot E_{\rm iso} t_{\rm w}) \ .
\end{equation}  
Some of the following results, however, do not rely on this assumption. 

\subsection{Processes involved}

\subsubsection{Collapse to Explosion}
The first function we have to consider is the dynamical effect 
(\S 3.1) of the wind to the collapsing stellar mantle; 
the overlying stellar mantle continues to collapse onto the central 
remnant, and the wind does not always have a sufficiently large momentum to 
overcome the ram pressure of the infalling materials (Fig. 1a). 
The importance of the dynamical effect to determine the outcome of a wind/jet-driven 
supernova explosion was pointed out by Maeda (2004) and numerically 
examined by Maeda (2004) and Tominaga et al. (2007b).
Some numerical calculations have included the dynamical effect self-consistently (e.g., 
Maeda \& Nomoto 2003; Maeda 2004; Tominaga et al. 2007b; Tominaga 2009), 
while others have not (e.g., Maeda et al. 2002; Nagataki et al. 2003). 

\subsubsection{Production of $^{56}$Ni}
Once the shock wave is launched, it propagates outward into the stellar mantle 
(Fig. 1b). The kinetic energy is now converted to the thermal energy 
following the shock wave propagation. 
The temperature behind the shock wave is initially so high 
that the nuclear burning converts the initial stellar composition (mostly 
oxygen and silicon) mainly to $^{56}$Ni (\S 3.2.1). 
The temperature decreases as the shock wave moves outward; once the temperatures drops 
below $\sim 5 \times 10^{9}$ K, then the efficiency of the production 
of $^{56}$Ni decreases rapidly. 

At the same time, the temperature of 
the injected wind itself may also be sufficiently high so that a fraction 
of this material may be 
converted to $^{56}$Ni (\S 3.2.2). In most of the past numerical studies, 
the production of $^{56}$Ni in shocked stellar mantle 
has been investigated in detail. 
The ejection of $^{56}$Ni in the wind has been examined by 
a different kind of numerical simulations involving the innermost 
part of the collapsing star (e.g., MacFadyen \& Woosley 1999; Nagataki et al. 2007). 
However, in such simulations, it is practically 
difficult to follow the dynamics of the 
shock wave propagating into the stellar mantle; Thus, it has not been yet clear 
how the amount of 
$^{56}$Ni synthesized in the wind is connected to properties of the 
progenitor star and of the resulting supernova. 

\subsubsection{Geometry of the ejecta}
Finally, the properties of the wind affect the shape of the supernova ejecta,
as the shock wave propagates through the stellar mantle. 
This has been directly examined in the past numerical studies, but restricted in 
the parameter space. In this paper, 
we derive a simple estimate of the shape, in terms of the properties of the wind 
(\S 3.3).

\section{Wind/jet-driven explosions}

\subsection{Collapse to Explosion}
In this section, we examine dynamical effect of the wind 
on the collapsing stellar mantle (Fig. 1a). 
We follow analysis similar to that given by Fryer \& M\'esz\'aros (2003) 
for a standard neutrino-driven delayed explosion for (canonical) supernovae.  

We evaluate outcome of the 
interaction between the wind and the infalling materials 
at radius $R_{\rm w} \sim 8 \times 10^7$ cm (for $M_{\rm ms} = 40M_{\odot}$) 
and $1.2 \times 10^8$ cm ($M_{\rm ms} = 25M_{\odot}$), 
where the presupernova enclosed mass ($M_r$) is $1.4M_{\odot}$. 
As time goes by, materials initially at larger $M_r$ collapses 
to the radius $R_{\rm w}$, and add to the mass of the central remnant ($M_{\rm BH}$). 
During a whole period before the launch 
of the shock, the temporal evolution of the system is thus specified by the 
central remnant mass $M_{\rm BH}$ which monotonically increases as a function of time 
(i.e., larger $M_{\rm BH}$ for later time). 

If the trajectory of the infalling materials is that of free fall, the
density ($\rho_{\rm m}$) and velocity ($v_{\rm m}$) of the infalling
materials at $R_{\rm w}$ are written as 
\begin{eqnarray}
\rho_{\rm m} & \sim & \rho_{\rm presn} 
\left(\frac{R_{\rm presn}}{R_{\rm w}}\right)^{3/2} \ , \\ 
v_{\rm m} & \sim & \left(\frac{2 G M_r}{R_{\rm w}}\right)^{1/2} \ . 
\end{eqnarray}  
The subscript "presn" is used for the pre-collapse initial 
values for the material at the mass coordinate $M_r$. 
The subscript "m" refers to quantities of the same material 
(at $M_r$) when it collapses to $R_{\rm w}$ (Figure 1a).  
Figure 2 shows snap shots of the density structure 
for different $M_{\rm BH}$ for $M_{\rm ms} = 40 M_{\odot}$. 

\begin{figure}
   \centering
   \includegraphics[width=0.45\textwidth]{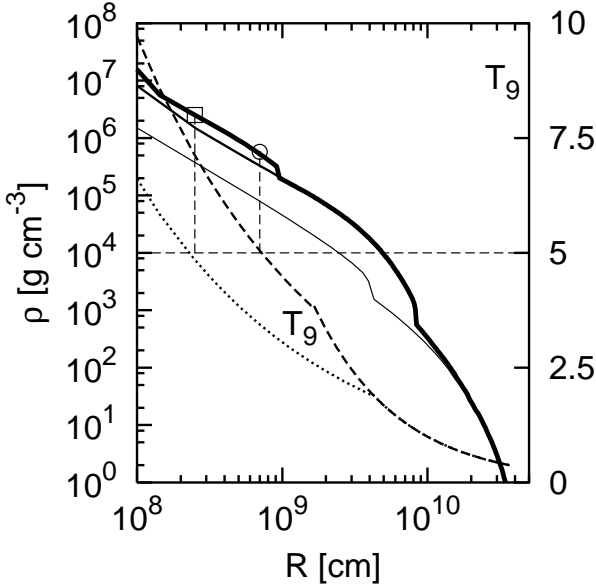}
      \caption{Density Structure (left vertical axis label) 
following the gravitational collapse, 
at $M_{\rm BH} = 2.4 M_{\odot}$ (thick-solid line), $3.4 M_{\odot}$ 
(medium-solid), and $13 M_{\odot}$ (thin-solid). Also shown is the 
postshock temperature ($T_{9} \equiv T/10^{9}$ K; right vertical 
axis label) as a function of the shock radius, for $M_{\rm BH} = 
2.4 M_{\odot}$, $E_{{\rm iso}, 51} = 30$, and 
$\dot E_{\rm iso} = 10$ (dashed line) or $1$ (dotted line). 
The progenitor model ($M_{\rm ms} = 40 M_{\odot}$) 
is from Nomoto \& Hashimoto (1988). 
The position within which the postshock temperature is above 
$5 \times 10^{9}$ K (horizontal thin-dashed line), 
for $M_{\rm BH} = 2.4 M_{\odot}$, 
is marked by the open circle ($\dot E_{{\rm iso}, 51} = 10$) 
and by the open square ($\dot E_{{\rm iso}, 51} = 1$). 
}
         \label{fig2}
\end{figure}

The outward shock wave is launched 
if the ram pressure of the wind 
($P_{\rm w}$) overcomes that of the infalling materials ($P_{\rm m}$), i.e., 
\begin{equation}
P_{\rm w} \sim {\Gamma_{\rm w}}^2 \rho_{\rm w} {v_{\rm w}}^2 
\ge P_{\rm m} = \rho_{\rm m} v_{\rm m}^2 
\equiv \rho_{\rm m} \bar g \frac{G M_{\rm BH}}{R_{\rm w}} \ .
\end{equation}
Here $M_{\rm BH}$ is the mass of the BH 
when the material at $M_r$ ($= M_{\rm BH}$) collapses to $R_{\rm w}$. 
Although a numerical constant $\bar g = 2$ in the above estimate, 
we find $\bar g = 1/2$ yields a better 
representation of a set of numerical simulations (Maeda 2004; 
Tominaga et al. 2007b). We assume $\bar g = 1/2$ throughout the paper. 

Equation (7) can be expressed in terms of $\dot E_{\rm w}$ and $\dot M_{\rm w}$, 
which are related to other properties of the wind/jet as follows. 
\begin{eqnarray} 
\dot E_{\rm w} & \sim & A_{\rm w} \Gamma_{\rm w} (\Gamma_{\rm w} - 1)
\rho_{\rm w} v_{\rm w} c^2 \ , {\rm and}\\ 
\dot M_{\rm w} & = & A_{\rm w} \Gamma_{\rm w} \rho_{\rm w} v_{\rm w} \ , 
\end{eqnarray} 
where $A_{\rm w}$ 
is the area subtended by the wind. 
The asymptotic expressions for the momentum balance are 
derived by substituting equations (8) and (9) into equation (7); 
\begin{eqnarray}
\sqrt{\dot E_{\rm w} \dot M_{\rm w}} & \equiv & 
f_{\Omega} \sqrt{\dot E_{\rm iso} \dot M_{\rm iso}} 
\ge \frac{1}{\sqrt{2}} 
A_{\rm w} \rho_{\rm m} \bar g \frac{G M_{\rm BH}}{R_{\rm w}} \nonumber\\
& & {\rm for} \ \Gamma_{\rm w} \sim 1 \ , \\
\dot E_{\rm w} & \equiv & f_{\Omega} \dot E_{\rm iso} 
\ge c A_{\rm w} \rho_{\rm m} \bar g \frac{G M_{\rm BH}}{R_{\rm w}} \nonumber\\ 
& & {\rm for} \ \Gamma_{\rm w} \gg 1 \ . 
\end{eqnarray} 

The RHS's of equations (10) and (11), except for $A_{\rm w}$ depending on $f_{\Omega}$, 
are completely determined by the progenitor structure, as a function of $M_{\rm BH}$ 
(Fig. 2). The requirement for $\dot E_{\rm w}$ and $\dot M_{\rm w}$ 
in terms of $\dot E_{\rm iso}$ and $\dot M_{\rm iso}$ 
is shown in Figure 3 as a function of $M_{\rm BH}$ (i.e., as a function of 
time). Hereafter, we use a notation $X_{n} \equiv X/10^{n}$ in CGS unit. 

\begin{figure}
   \centering
   \includegraphics[width=0.4\textwidth]{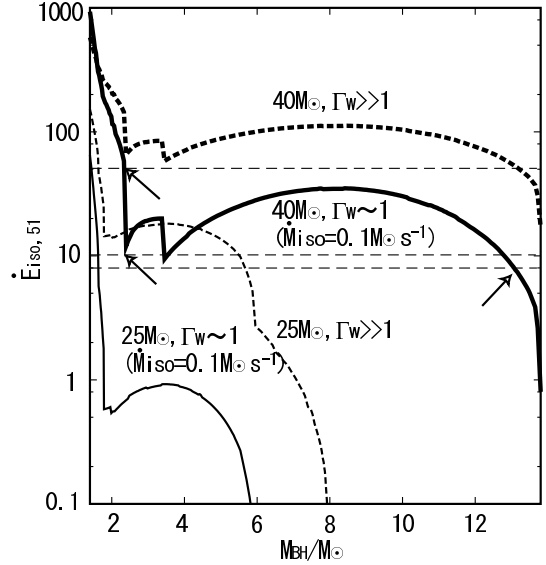}
      \caption{The requirement for $\dot E_{\rm iso}$ 
to initiate the explosion as a function of $M_{\rm BH}$. 
For $\Gamma_{\rm w} \sim 1$, the case with 
$\dot M_{\rm iso} = 0.1M_{\odot}$ s$^{-1}$ is shown for presentation  
(note that the product $\dot E_{\rm iso} \dot M_{\rm iso}$ should exceed 
the value corresponding $P_{\rm m}$ if $\Gamma_{\rm w} \sim 1$; equation 10).
Three horizontal dashed lines at $\dot E_{{\rm iso}, 51} = 50$, $10$, and $7$ 
are shown for illustrating purpose (arrows indicate the position where 
the outward shock wave is launched); 
Let us assume $M_{\rm ms} = 40 M_{\odot}$, $\Gamma_{\rm w} \sim 1$ and $\dot M_{\rm iso} = 
0.1 M_{\odot}$ s$^{-1}$. If $\dot E_{{\rm iso}, 51} = 50$ or $10$,  
the outward shock wave is launched at the intersections between the Si- and 
CO-layers at $M_{\rm BH} \sim 2.4 M_{\odot}$. On the other hand, if 
$\dot E_{{\rm iso}, 51} = 7$, the momentum can not exceed the infalling 
ram pressure until nearly the whole CO star collapses to the central remnant 
at $M_{\rm BH} \sim 13 M_{\odot}$. 
              }
         \label{fig3}
\end{figure}

If $\Gamma_{\rm w} \sim 1$, it is necessary that the 
product $\dot E_{\rm iso} \dot M_{\rm iso}$ 
should be larger than a specific value given by equation (10), which 
corresponds to $P_{\rm m}$ as a function of $M_{\rm BH}$, in order to initiate 
the explosion. For $\Gamma_{\rm w} \gg 1$, $\dot E_{\rm iso}$ should exceed the value 
given by equation (11), and the requirement becomes independent from $\dot M_{\rm iso}$. 
In other ward, once the temporal evolution of $\dot E_{\rm iso}$ and $\dot M_{\rm iso}$ is given, 
the outward shock wave is launched when $\dot E_{\rm iso} \dot M_{\rm iso}$ 
(or $\dot E_{\rm iso}$ if $\Gamma_{\rm w} \gg 1$) is higher than the
specific value representing $P_{\rm m}$ (Fig. 3). 

Although $P_{\rm m}$ overall decreases as a function of $M_{\rm BH}$ 
following the density decrease, 
jumps are seen at the edges of 
the characteristic hydrostatic burning layers. 
Assuming freefall, $P_{\rm m} \propto \rho_{\rm presn} 
R_{\rm presn}^{3/2} M_{\rm BH}$. Within each layer, 
$\rho_{\rm presn}$ drops slowly as a function of $M_{\rm BH}$ and $R_{\rm presn}$, 
making $P_{\rm m}$ nearly constant as a function of $M_{\rm BH}$. 
On the other hand, at the edges, 
$\rho_{\rm presn}$ decreases suddenly, leading to rapidly decreasing $P_{\rm m}$. 
As a result, local minimum values appear at the intersection of different layers (Fig. 3). 
It infers that the explosion is likely initiated at one of these intersections, 
but not within each layer. 

This behavior of $P_{\rm m}$ provides an interesting implication. 
For illustration purpose, 
let us take the case with $M_{\rm ms} = 40M_{\odot}$ 
and $\Gamma_{\rm w} \sim 1$, 
assuming that $\dot E_{{\rm iso}, 51}$ and $\dot M_{\rm iso}$ are 
constant in time. 
If $\dot E_{{\rm iso}, 51} \dot M_{{\rm iso}, \odot} \ga 1$, 
then the explosion is initiated below the Fe/Si interface ($M_{\rm BH} < 2.4M_{\odot}$). 
If $0.8<\dot E_{{\rm iso}, 51} \dot M_{{\rm iso}, \odot} < 1$, the explosion 
position jumps to the Si/CO interface ($M_{\rm BH} \sim 3.4M_{\odot}$). 
Then, if $\dot E_{{\rm iso}, 51} \dot M_{{\rm iso}, \odot} \la 0.8$, almost whole CO layer collapses onto the central remnant. 
Thus, we conclude that a small difference of the wind properties can 
lead to totally different outcome, with the critical value 
$\dot E_{{\rm iso}, 51} \dot M_{{\rm iso}, \odot} \sim 0.8 - 1$ 
(for $\Gamma_{\rm w} \sim 1$) or $\dot E_{{\rm iso}, 51} \sim 70$ 
(for $\Gamma_{\rm w} \gg 1$). 
The critical value depends on $M_{\rm ms}$, and 
is smaller for a less massive progenitor 
(Fig. 3).

\subsection{Production of $^{56}$Ni}
Once the ram pressure of the wind ($P_{\rm w}$) 
overcomes the ram pressure of the infalling material 
($P_{\rm m}$), an outward shock wave is launched and explosive
nucleosynthesis takes place. 
In this section, we examine production of $^{56}$Ni 
as the shock wave passes through the stellar mantle (Fig. 1b). 
We first discuss production of $^{56}$Ni in 
a stellar mantle heated by the shock wave (\S 3.2.1), 
then we comment on production of $^{56}$Ni within the wind/jet (\S 3.2.2). 

\subsubsection{Shocked stellar mantle}

The shocked mantle can become the predominant site for 
the $^{56}$Ni synthesis, if the shock wave sweeps up a 
large amount of the stellar mantle, at least 
$\sim 0.1 M_{\odot}$. This inevitably results in 
a non-relativistic shock wave, 
even if the wind/jet at its emergence from the central system 
at $R_{\rm w}$ is highly relativistic. 
Let us denote the average velocity and the mass 
of the expanding material swept up by the shock wave, by 
$\Gamma_{\rm shock}$ and $M_{\rm shock}$, respectively, then 
\begin{equation}
\Gamma_{\rm shock} \sim 1 + 6 \times 10^{-4} 
\frac{\dot E_{{\rm w}, 51} t}
{M_{{\rm shock}, \odot}} \ , 
\end{equation} 
at time $t$ (in second). In order to accelerate 
$\sim 0.1 M_{\odot}$ of the stellar mantle materials 
to $\Gamma_{\rm shock}=100$,
the explosion energy is required to be higher than $10^{55}$ ergs

We can thus use non-relativistic approximation for 
the postshock temperature ($T$) as a function of the radius of the shock wave 
($R_{\rm sw}$). For a radiation-dominated 
fireball expanding into a uniform medium with the density $\bar \rho$ 
(Maeda \& Nomoto 2003), 
\begin{equation}
  T_9 \sim \left\{ 
     \begin{array}{ll}
        5.7 \dot E_{{\rm iso}, 51}^{1/6} R_{{\rm sw}, 8}^{-1/3} {\bar \rho_{6}}^{1/12} 
         & \quad \mbox{for $t<t_{\rm w}$} \ , \\
        24 E_{{\rm iso}, 52}^{1/4} R_{{\rm sw}, 8}^{-3/4} & \quad \mbox{for $t \ge t_{\rm w}$} \ . 
     \end{array}\right.
\end{equation}
This expression explicitly utilizes the assumption that 
$\dot E_{\rm iso}$ is constant in time for $t \le t_{\rm w}$ 
and zero afterward.

\begin{figure}
   \centering
   \includegraphics[width=0.4\textwidth]{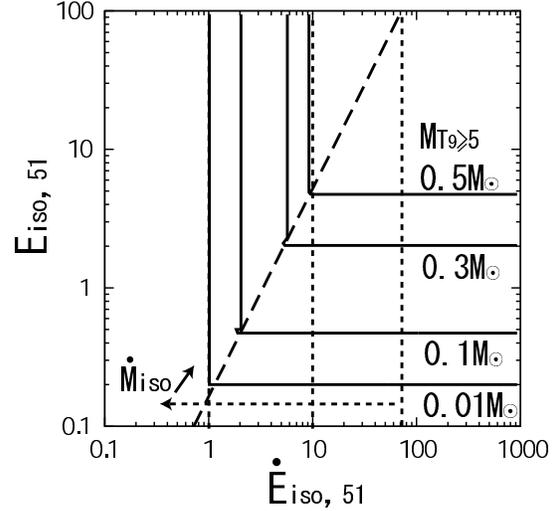}
      \caption{The isotropic 
mass of $^{56}$Ni synthesized in the shocked stellar mantle 
($M_{\rm T_9\ge5}$; solid contours), 
for $M_{\rm BH} =2.4M_{\odot}$ and $M_{\rm ms} = 40M_{\odot}$ 
($13.8M_{\odot}$ CO star). 
The dashed aslant line shows $\dot E_{{\rm iso}, 51} = 
\dot E_{{t_{\rm w}, 51}}$ ($E_{\rm iso}$) as is defined by equation (14). 
Note that the possible value of $\dot E_{\rm iso}$ 
that leads to $M_{\rm BH} = 2.4 M_{\odot}$ is a function of $\dot M_{\rm iso}$ 
(see the main text). 
Three cases are shown for illustration, assuming the 
temporally constant wind injection: If $\Gamma \gg 1$, the required 
value for $\dot E_{\rm iso}$ is independent from $\dot M_{\rm iso}$, and 
it is required that $\dot E_{{\rm iso}, 51} \sim 70$ (otherwise the outward shockwave 
cannot be launched at $M_{\rm BH} = 2.4 M_{\odot}$.) 
If $\Gamma_{\rm w} \sim 1$, it is required that the product $\dot E_{\rm iso} \dot M_{\rm iso}$ 
should exceed a certain value, i.e., $\dot E_{\rm iso} \dot M_{\rm iso} \sim 1$ 
for the outward shock wave being launched at $M_{\rm BH} = 2.4 M_{\odot}$ 
(e.g., $\dot E_{{\rm iso}, 51} \sim 10$ if $\dot M_{{\rm iso}, \odot} = 0.1$, 
and $\dot E_{{\rm iso}, 51} \sim 1$ if $\dot M_{{\rm iso}, \odot} \sim 1$). 
         }
         \label{fig4}
\end{figure}

Since $M_{\rm BH}$ increases monotonically with time, the density
structure of the stellar mantle is specified for given $M_{\rm BH}$ (Fig. 2). 
Therefore, we can estimate the enclosed mass within a sphere in which 
the post-shock materials attain a certain temperature 
(Fig. 2; see also figure 7 of Maeda \& Nomoto 2003). 
$^{56}$Ni is synthesized in the region ($M_{\rm T_9\ge5}$) where $T_9 \ge 5$.
Figure 4 shows $M_{\rm T_9\ge5}$ [$= M$($^{56}$Ni) $/f_{\Omega}$; the isotropic 
equivalent value for $M$($^{56}$Ni)] 
for $M_{\rm BH} = 2.4M_{\odot}$ and $M_{\rm ms} = 40M_{\odot}$. 

$M_{\rm T_9\ge5}$ depends not only on $E_{\rm iso}$ and $\dot E_{\rm iso}$, 
but also on $M_{\rm BH}$.
On the other hand, as concluded in \S 3.1, there is a condition in terms of 
$\dot E_{\rm iso}$ and $\dot M_{\rm iso}$, to initiate the explosion at given $M_{\rm BH}$: 
This indirectly relates $\dot M_{\rm iso}$ to $M_{\rm BH}$. 
In order to obtain $M_{\rm BH} = 2.4 M_{\odot}$ and $M_{\rm ms} = 40 M_{\odot}$, 
it is required that $\dot E_{{\rm iso}, 51} \dot M_{{\rm iso}, \odot} \ga 1$ 
(for $\Gamma_{\rm w} \sim 1$) or that $\dot E_{{\rm iso}, 51} \ga 70$ 
(for $\Gamma_{\rm w} \gg 1$). 
Vertical dotted lines in Figure 4 explicitly describe the requirement: 
(1) for $\Gamma \gg 1$, $\dot E_{{\rm iso}, 51} = 70$, 
(2) for $\Gamma \sim 1$ and $\dot M_{{\rm iso}, \odot} = 0.1$, 
$\dot E_{{\rm iso}, 51} = 10$, and 
(3) for $\Gamma \sim 1$ and $\dot M_{{\rm iso}, \odot} = 1$, 
 $\dot E_{{\rm iso}, 51} = 1$. 

Interestingly, the dependence of $M$($^{56}$Ni) on $\dot E_{\rm iso}$ 
and $E_{\rm iso}$ (Fig. 4) changes the behavior at a specific 
aslant line in the $\dot E_{\rm iso} - E_{\rm iso}$ plane. 
The line is obtained by equalizing the two RHS's of equations (13) 
and defined as
\begin{equation}
\dot E_{{t_{\rm w}}, 51} (E_{\rm iso}) 
\equiv \left(\frac{2.9}{T_9}\right)^{-10/3} {\bar \rho_6}^{-1/2} 
{E_{{\rm iso}, 51}}^{2/3} \ ,
\end{equation}
with $T_9 = 5$, i.e., the temperature necessary for the production 
of $^{56}$Ni. For given $E_{\rm iso}$, $M$($^{56}$Ni) is larger for 
larger $\dot E_{\rm iso}$ as long as $\dot E_{\rm iso} \le \dot E_{t_{\rm w}}$, 
but saturated for $\dot E_{\rm iso} > \dot E_{t_{\rm w}}$. 

The dividing energy injection rate, $\dot E_{t_{\rm w}} (E_{\rm iso})$, 
appears, because of the different behavior of the post-shock temperature 
before and after $t = t_{\rm w}$ (Maeda 2004). 
The mass of $^{56}$Ni is determined by the radius of the outward shock wave 
when the temperature drops down to $T_{9} = 5$. 
If $\dot E_{\rm iso} \le \dot E_{t_{\rm w}}$ (for given $E_{\rm iso}$), 
then the temperature behind the shock wave drops down to $T_{9} = 5$ 
during the phase when the energy source is still active (i.e., 
$t < t_{\rm w}$). Thus, larger $\dot E_{\rm iso}$ corresponds to 
the larger amount of energy contained behind the shock wave, 
leading to the larger radius of the shock wave when the condition 
$T_{9} = 5$ is satisfied (e.g., see Fig. 2). 
On the other hand, the condition 
$\dot E_{\rm iso} > \dot E_{t_{\rm w}}$ corresponds to the 
case in which all the energy from the central source is liberated 
in short time scale such that the temperature is still high 
($T_{9} > 5$) at $t = t_{\rm w}$. The following evolution of 
the shock wave, which determines $M_{\rm T_9\ge5}$, 
is controlled by $E_{\rm iso}$, not by $\dot E_{\rm iso}$. 

It has been suggested that $M$($^{56}$Ni) is larger for larger $E_{\rm iso}$ 
(e.g., Nakamura et al. 2001b), but this holds true 
only if the energy is generated and liberated almost 
promptly (i.e., $t_{\rm w} \to 0$ and $\dot E_{\rm iso} \to \infty$). 
For example, if $\dot E_{{\rm iso}, 51} = 1$, then 
$M$($^{56}$Ni) does not depend on the total energy $E_{\rm iso}$ (or $E_{\rm w}$) 
(as long as $E_{{\rm iso}, 51} \ga 0.2$) (Fig. 4).

\subsubsection{$^{56}$Ni in the wind/jet}

Mass of the wind is 
\begin{equation}
M_{\rm w} = \frac{\dot M_{\rm w}}{\dot E_{\rm w}} E_{\rm w} 
\sim \frac{E_{\rm w}}{c^2 (\Gamma_{\rm w} - 1)} \ ,
\end{equation}
which must be smaller than the mass of the accreted 
materials, i.e., $M_{\rm w} < M_{\rm BH} - 1.4M_{\odot}$. 
Composition of the materials within the disk wind is largely uncertain, 
because it depends on the thermal history of the wind 
in the vicinity of the central remnant (i.e., well below $R_{\rm w}$). 
The deep understanding of the composition of the wind 
material requires numerical calculations or analytic investigation 
of the innermost part of the collapsing star (e.g., MacFadyen \& Woosley 1999; 
Pruet, Thompson, \& Hoffman 2004; Nagataki et al. 2007). 

However, it is still possible to make a rough estimate, in order to 
see in what regions in the $\dot E_{\rm iso} - \dot M_{\rm iso}$ plane 
the disk wind is important. 
Assuming that the energy of the wind is initially thermal energy-dominated 
near the central remnant, the typical entropy of the wind material 
$s \equiv S/(k_{\rm B}/m_{\rm u})$ 
(where $k_{\rm B}$ and $m_{\rm u}$ are the Boltzmann constant and 
atomic unit mass, respectively) is written by 
$\dot E_{\rm iso}$ and $\dot M_{\rm iso}$ as follows: 
\begin{equation}
  s \sim \left\{ 
     \begin{array}{ll}
        22 (\dot E_{{\rm iso}, 51} \dot M_{{\rm iso}, \odot})^{3/8} 
         & \quad \mbox{for $\Gamma_{\rm w} \sim 1$} \ , \\
        11 \dot E_{{\rm iso}, 52}^{3/4} 
         & \quad \mbox{for $\Gamma_{\rm w} \gg 1$} \ . 
     \end{array}\right.
\end{equation} 

We find that $s \gg 1$ in the parameter range 
of our interest; $s$ can be as small as about unity, 
only when either $\dot E_{\rm iso} \dot M_{\rm iso}$ or $\dot E_{\rm iso}$ 
is very small, but in such a case the whole star collapses onto the central remnant (\S 3.1). 
Thus, when $\dot E_{\rm iso}$ and $\dot M_{\rm iso}$ are large enough to
result in a supernova explosion, the wind material 
should experience the strong $\alpha$-rich freezeout, leaving mainly $^{4}$He, 
not $^{56}$Ni. The strong $\alpha$-rich freezeout is consistent 
with numerical calculations of the collapsing star (Nagataki et al. 2007). 
In what follows, we take $M$($^{56}$Ni) $ = 0.2 M_{\rm w}$ in the wind
which is typical for the strong $\alpha$-rich freezeout 
(e.g., Pruet et al. 2004; 
but see \S 5 for further discussion on the uncertainty).

\subsection{Geometry of the ejecta}

The geometry of the bulk supernova materials provide 
a useful constraint on the model, since recently more and 
more observational data have been available to 
address the ejecta geometry (see Wang \& Wheeler 2008 for a review of 
spectropolarimetry; 
for recent spectroscopy, see Maeda et al. 2008; Modjaz et al. 2008). 
Here we give a rough estimate of how the 
geometry depends on the wind 
parameters. Note that the geometry of the bulk supernova materials 
is different from the geometry of the wind at the injection 
measured by $f_{\Omega}$; Even if the wind is initially collimated 
($f_{\Omega} < 1$), the bulk expansion of the stellar mantle,  
as is induced by the wind, can be 
less collimated or even quasi-spherical. 

The shock breakout time ($t_{\rm sb}$) of the wind/jet 
is estimated by 
\begin{eqnarray}
t_{\rm sb} & \sim & R_{\rm CO}/v_{\rm shock} \ ,\\
v_{\rm shock} & \equiv &
\left(\frac{2 \dot E_{\rm iso} t_{\rm sb}}{(M_{\rm CO} - M_{\rm BH})}\right)^{1/2} \ . 
\end{eqnarray}
Here $R_{\rm CO}$ and $M_{\rm CO}$ are the radius and the mass of the 
CO star. 
For $M_{\rm ms} = 40M_{\odot}$, these values are $R_{\rm CO} \sim 10^{10}$ cm 
and $M_{\rm CO} \sim 13.8M_{\odot}$. 

As the wind/jet pushes the stellar mantle, the wind/jet loses 
its energy by depositing its energy into the surroundings. 
If the energy injection is terminated before 
it breaks through the progenitor surface ($R_{\rm CO}$), 
a large fraction of the energy of the wind/jet are transferred 
to the stellar mantle. This results in a quasi-spherical explosion 
of the bulk of the stellar mantle, even if the wind/jet is initially collimated.  
This happens if the following condition is satisfied. 
\begin{equation}
t_{\rm w} \sim \frac{E_{\rm iso}}{\dot E_{\rm iso}} < t_{\rm sb} \sim 
\left(\frac{{R_{\rm CO}^2} (M_{\rm CO} - M_{\rm BH})}{2 \dot E_{\rm iso}}\right)^{1/3} \ .
\end{equation}
On the other hand, if $t_{\rm w} > t_{\rm sb}$, then 
a large amount of materials are ejected toward the jet 
direction as compared to the equatorial direction, resulting in a 
strongly jetted explosion.

\section{Comparison to Observations}
The expected SN properties can be expressed by 
$\dot E_{\rm iso}$ and $\dot M_{\rm iso}$ and the temporal evolution 
of these, once the other parameters are 
specified ($M_{\rm ms}$, $E_{\rm iso}$ or $E_{\rm w}$, and $f_{\Omega}$). 
For comparison to the observation presented below, 
we examine the simplest case, in which $\dot E_{\rm iso}$ and 
$\dot M_{\rm iso}$ are constant in time for $t < t_{\rm w}$ and 
zero afterward (i.e., $E_{\rm iso} = \dot E_{\rm iso} t_{\rm w}$). 
This is a situation examined in most of the previous numerical studies, 
except for MacFadyen et al. (2001) and 
Maeda \& Nomoto (2003) who examined the case where the energy injected 
by the jet (wind) is connected to the accretion rate to the central
remnant. 

First, the momentum balance determines when the outward shock wave is launched, 
yielding $M_{\rm BH}$ at this time as a function of $\dot E_{\rm iso}$ and $\dot M_{\rm iso}$ 
(\S 3.1; Figure 3). Then, the analysis presented in 
\S 3.2.1 gives us the mass of $^{56}$Ni, as this is given as a function of $M_{\rm BH}$, 
$\dot E_{\rm iso}$, and $E_{\rm iso}$ (Figure 4). The mass of $^{56}$Ni in the disk 
can be roughly evaluated using the result of \S 3.2.2 (equation 15). Finally, 
the typical geometrical feature can be derived as a function of 
$\dot E_{\rm iso}$, $E_{\rm iso}$, $M_{\rm BH}$ (\S 3.3; equation 19). 

Using the expressions derived in \S 3, 
we characterize properties of a SN explosion 
as a function of $\dot E_{\rm iso}$ and $\dot M_{\rm iso}$ 
for a given progenitor model ($M_{\rm ms}$). 
Other parameters are $f_{\Omega}$, 
$E_{\rm w}$ (or $t_{\rm w}$), but these can be 
set without large ambiguity by observations for 
some GRB-SNe of special interest. 

\subsection{SN 1998bw: the origin of $^{56}$Ni}

SN Ic 1998bw is a prototypical SN associated with a GRB. 
For SN 1998bw, intensive observational data in the optical 
wavelength are available: 
modeling these observations 
(by one-dimensional radiation transfer calculations; 
Iwamoto et al. 1998; Woosley et al. 1999; Nakamura et al. 2001a) 
suggests that $E_{\rm iso} \sim 30$ and $M_{\rm ms} 
\sim 40 M_{\odot}$. 
Adding to this, there exists intensive study for this object 
using multi-dimensional radiation transfer calculations 
(Maeda 2006a; Maeda et al. 2006bc; Tanaka et al. 2007). 
The study suggests that the intrinsic explosion energy is smaller than 
the isotropic value by a factor of about 3, inferring that 
$f_{\Omega} \sim 0.3$ is a good approximation. 

Figure 5a shows the result for $E_{{\rm iso}, 51} = 30$, 
$f_{\Omega} = 0.3$, and $M_{\rm ms} = 40M_{\odot}$. 
Here, we try to constrain the properties of the wind 
generated by the activity of the central engine of SN 1998bw, i.e., 
$\dot E_{\rm iso}$ and $\dot M_{\rm iso}$ (or equivalently, $\dot E_{\rm w}$ and 
$\dot M_{\rm w}$). 
For SN 1998bw, observations constrain three quantities, the ejecta
mass, $M$($^{56}$Ni), and the shape of the ejecta: 
(1) The ejecta contain a large amount of the CO-core materials, 
so that $M_{\rm BH} < 10M_{\odot}$, as inferred by the optical light curve 
and spectra (\S 3.1). 
(2) $M$($^{56}$Ni) $\sim 0.4M_{\odot}$ to explain its peak luminosity (\S 3.2). 
(3) The ejecta are suggested to be aspherical, but still a large amount 
of materials are ejected into the equatorial direction as inferred especially 
by spectra at $\sim 1$ year since the explosion 
(Patat et al. 2001; Mazzali et al. 2001; Maeda et al. 2002). 
This indicates that the expanding supernova ejecta are quasi-spherical, 
rather than extremely bipolar (\S 3.3)

The three conditions are simultaneously satisfied 
if $\dot E_{{\rm iso}, 51} \dot M_{{\rm iso}, \odot} \ga 1$  
and $\dot E_{{\rm iso}, 51} \ga 20$ (for $\Gamma_{\rm w} \sim 1$), or  
$\dot E_{{\rm iso}, 51} > 70$ (for $\Gamma_{\rm w} \gg 1$) 
(region A in Figure 5b). 
In these cases, $^{56}$Ni is predominantly 
produced at the shocked stellar mantles. 
The wind origin for $^{56}$Ni is disfavored for SN 1998bw. 
The wind contribution exceeds the shocked mantle contribution only if 
$\dot E_{{\rm iso}, 51} \la 5$ and 
$\dot M_{{\rm iso}, \odot} \ga 0.2$ (i.e., region C of Figure 5b). 
However, this combination of the parameters results in $t_{\rm w}/t_{\rm sb} > 1$ 
and an essentially bipolar explosion, as is inconsistent with the observation. 

In short, the wind should be massive and long-lived (i.e., large $\dot M_{\rm iso}$ 
and large $t_{\rm w}$; see equation 15), 
in order to produce a large amount of $^{56}$Ni within the wind 
(as large as $\sim 0.4 M_{\odot}$). However, such an explosion 
with a long-lived energy injection results in the extremely bipolar 
explosion, since the outward shock wave can reach the stellar surface 
before the energy injection is terminated: This argues against the 
wind-origin of $^{56}$Ni in SN 1998bw.

\begin{figure*}
   \centering
   \includegraphics{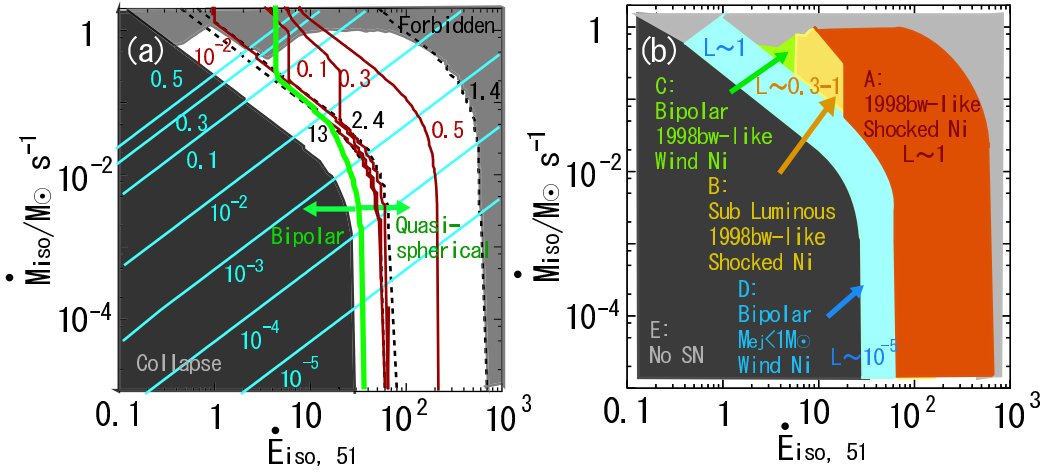}
   \caption{(a) Estimate of $M$($^{56}$Ni) in the $\dot E_{\rm iso} - \dot M_{\rm iso}$ plane. 
The progenitor is the $13.8M_{\odot}$ CO star ($M_{\rm ms} = 40M_{\odot}$), taken 
from Nomoto \& Hashimoto (1988). 
The other two parameters are set as $E_{{\rm iso}, 51} = 30$ and $f_{\Omega} = 0.3$, 
representing the bright GRB-SN 1998bw. 
$M$($^{56}$Ni)/$M_{\odot}$ is shown for the shocked stellar mantle (red contours) 
and the wind (blue contours). 
$M_{\rm BH}$/$M_{\odot}$ is shown by the black-dotted contours. 
The shaded region denoted by "Forbidden" is excluded 
by the requirement that the ejected mass must not exceed the 
available accreted mass budget. The dark shaded region denoted by "Collapse" 
corresponds to whole collapse of the CO star. 
The green curve shows the line where $t_{\rm w} \sim t_{\rm sb}$. 
(b) Expected characteristics of GRB-SNe for $M_{\rm ms} = 40M_{\odot}$. 
In each region, "L" denote a rough estimate of the SN peak luminosity 
normalized by that of SN 1998bw.}
   \label{fig5}
\end{figure*}

\subsection{Diversity of GRB-SNe}
Figure 5b shows the expected characteristics of GRB-SNe for 
$E_{\rm iso} = 30$, $f_{\Omega} = 0.3$, and $M_{\rm ms} = 40M_{\odot}$. 
A variety of features are predicted for the wind-driven supernovae 
depending on $\dot E_{\rm iso}$ and $\dot M_{\rm iso}$. 

If $\dot E_{{\rm iso}, 51} \dot M_{{\rm iso}, \odot} \ga 1$ ($\Gamma_{\rm w} \sim 1$)
or $\dot E_{{\rm iso}, 51} \ga 70$ ($\Gamma_{\rm w} \gg 1$), 
then the resulting SNe should be similar 
to SN 1998bw in the ejected mass. 
The features are further divided as follows (regions A, B, and C in 
Figure 5b): 
\begin{description}
\item {\bf (A) SN 1998bw-like -- GRB-SNe 1998bw, 2003dh, and 2003lw: }
In the parameter region A, the wind-driven supernova is similar to 
GRB-SN 1998bw, in virtually all the observed characteristics. 
The luminosity is similar to 
that of SN 1998bw, since $M$($^{56}$Ni) $\sim 0.3 - 0.6M_{\odot}$. 

This category can account for GRB-SNe 1998bw, 2003dh, and 2003lw. 
The region is relatively large in terms of $\dot E_{\rm iso}$ and 
$\dot M_{\rm iso}$, and the features of supernovae are insensitive to 
$\dot E_{\rm iso}$ and $\dot M_{\rm iso}$; 
This can explain why these three GRB-SNe are 
similar in the optical properties. 

\item {\bf (B) Sub-luminous SN 1998bw-like -- GRBs 040924 and 041006?: }
In the parameter region B, the expected supernovae are similar to SN 1998bw in the ejected mass 
and the geometry, but the difference is seen in $M$($^{56}$Ni). This indicates 
that the supernovae has similar optical properties 
with SN 1998bw, with the diversity in the luminosity, covering 
$\sim 0.3 - 1$ times that of SN 1998bw. This diversity arises 
because, in this parameter region, 
$\dot E_{\rm iso}$ is smaller than $\dot E_{t_{\rm w}}$, and 
thus $M$($^{56}$Ni) is dependent on $\dot E_{\rm iso}$ (\S 3.2.1) 
[note that $\dot E_{\rm iso}$ is larger than $\dot E_{t_{\rm w}}$ in region A, 
and thus $M$($^{56}$Ni) is independent from $\dot E_{\rm iso}$ unlike region B]. 
This parameter region B can account for 
sub-luminous SNe found as bumps in optical afterglows 
of some GRBs (e.g., GRBs 040924 and 041006).

\item {\bf (C) Bipolar SN 1998bw-like: }
In the parameter region C, 
a resulting SN is similar to SN 1998bw in its luminosity and ejecta mass, 
but the ejecta are more highly beamed than SN 1998bw. We have never directly 
observed such peculiar GRB-SNe. 
If such a GRB-SN is observed in future, 
it will provide a strong evidence of the diverse property of 
the central engine, in terms of $\dot E_{\rm iso}$ and $\dot M_{\rm iso}$. 

Here, $^{56}$Ni is mainly produced within the wind materials. 
The bipolar explosion results from the central energy source 
being active for a long time period (c.f., equation 19; e.g., 
low $\dot E_{\rm iso}$ smaller than $\dot E_{\rm w}$). 
As such, (1) the mass of $^{56}$Ni produced within the shocked stellar mantle 
is small, and (2) the wind should be massive to initiate the explosion 
(to provide the sufficiently large momentum in the non-relativistic regime). 
As a result, the wind contribution dominates over the shocked stellar mantle in 
the production of $^{56}$Ni for the (extremely) bipolar explosion case.

\end{description}

If $\dot E_{{\rm iso}, 51} \dot M_{{\rm iso}, \odot} \la 1$ ($\Gamma_{\rm w} \sim 1$)
or $\dot E_{{\rm iso}, 51} \la 70$ ($\Gamma_{\rm w} \gg 1$), 
then the resulting SNe should intrinsically different from 
SN 1998bw in the ejected mass (\S 3.1; regions D and E in Figure 5b). At the low end of 
$\dot E_{\rm iso}$ or $\dot M_{\rm iso}$, a supernova can 
never be triggered.

\begin{description} 

\item {\bf (D) Bipolar SN with the ejected mass $\la 1 M_{\odot}$ -- GRBs 060505 and 060614?: }
In the parameter region D, only a small amount of materials ($< 1M_{\odot}$) 
near the surface of the CO layer are ejected. 
The supernova ejecta are essentially bipolar in this case. 
$^{56}$Ni is mainly originated in the wind. 
Thus, $M$($^{56}$Ni) $\propto \dot M_{\rm w}/\dot E_{\rm w}$, ranging from 
$M$($^{56}$Ni) $< 10^{-5}M_{\odot}$ to $\sim 0.3M_{\odot}$. 
The expected luminosity is thus diverse, depending on $\dot E_{\rm iso}$ 
and $\dot M_{\rm iso}$. 
This may correspond to non-detection of supernova features in 
GRBs 060505 and 060614 (which may also be explained by region E below).

\item {\bf (E) No supernova -- GRBs 060505 and 060614?: }
In the parameter region E, the wind injection can never set the outward shock wave. 
This case corresponds to whole collapse of the progenitor CO star without 
a SN. 

\end{description} 

\section{CONCLUSIONS AND DISCUSSION}

In this paper, we discussed theoretically 
expected characteristics of a supernova driven by the wind/jet. 
We found that the resulting supernova features can be categorized 
as a function of $\dot E_{\rm iso}$ and $\dot M_{\rm iso}$. 
Thus, it is possible to constrain the nature of the central engine of GRBs 
by observations of associated supernovae. 

Results of the past numerical studies can be understood as a limiting case. 
For example, Nakamura et al. (2001b) showed that $M$($^{56}$Ni) is 
larger for larger $E_{\rm iso}$. In this paper, we have clarified that 
it holds true only for the limiting case where $t_{\rm w} \to 0$ and 
thus $\dot E_{\rm iso} \to \infty$. 
For an another example, 
Tominaga et al. (2007b) showed that 
$\dot E_{\rm iso}$ determines $M$($^{56}$Ni). In this paper, 
we have shown that this behavior appears as a limiting case where 
$\Gamma \gg 1$, and the dependence is different when $\Gamma \sim 1$. 

For SN 1998bw, we find that observations are reproduced only if 
$\dot E_{{\rm iso}, 51} \dot M_{{\rm iso}, \odot} \ga 1$ 
and $\dot E_{{\rm iso}, 51} \ga 20$ (if $\Gamma_{\rm w} \sim 1$), or  
$\dot E_{{\rm iso}, 51} \ga 70$ (if $\Gamma_{\rm w} \gg 1$) (region A in Fig. 5
b). 
We favor the shocked stellar mantle as the main site of the production of 
$^{56}$Ni. 

Furthermore, the observed diversity of supernovae associated (or not associated) with GRBs 
can be accounted for by the diverse properties of the wind from the central remnant 
($\dot E_{\rm iso}$ and $\dot M_{\rm iso}$). 
It is shown that the different wind properties 
can potentially explain the diversity of supernovae 
associated with GRBs. 

Further diversity can arise from different progenitors ($M_{\rm ms}$). 
Irrespective of $M_{\rm ms}$, the expected features of the wind-driven supernovae 
can be categorized as we did for $M_{\rm ms} = 40 M_{\odot}$ (\S 4), i.e., 
regions A -- E. 
The positions of the boundaries between different regions, as well as 
maximum $M$($^{56}$Ni) at the high end of $\dot E_{\rm iso}$ and $\dot M_{\rm iso}$, 
are dependent on $M_{\rm ms}$. 
For $M_{\rm ms} = 25M_{\odot}$, the boundary between regions A and D 
is located at $\dot E_{{\rm iso}, 51} \dot M_{{\rm iso}, \odot} 
\sim 0.05$ (for $\Gamma_{\rm w} \sim 1$) or $\dot E_{{\rm iso}, 51} \sim 20$ 
(for $\Gamma_{\rm w} \gg 1$), 
smaller than for $M_{\rm ms} = 40M_{\odot}$ by about one order of magnitude (Fig. 3). 
Using $f_{\Omega} = 0.3$ and $E_{{\rm iso}, 51} = 30$,  
maximum $M$($^{56}$Ni) 
is $\sim 0.24M_{\odot}$ even if $M_{\rm BH} = 1.4M_{\odot}$. 
Thus, $M_{\rm ms} = 25M_{\odot}$ never yield bright SN 1998bw-like SNe. 
We therefore confirmed that the prototypical GRB-SN 1998bw and the similar SNe 
2003dh and 2003lw must have originated from a massive progenitor ($M_{\rm ms} \sim 40 M_{\odot}$) 
from the nucleosynthesis argument.

Also, the diversity arising from $M_{\rm ms}$ may be 
important to explain the sub-luminous supernovae 
possibly associated with GRBs 040924 and 041006. 
These can be explained by $M_{\rm ms} \sim 40 M_{\odot}$
with smaller $\dot E_{\rm iso}$ and $\dot M_{\rm iso}$ than for 
SN 1998bw. Alternatively, it is also possible that 
the central engine provides $\dot E_{\rm iso}$ and 
$\dot M_{\rm iso}$ similar to SN 1998bw, but the progenitor mass is smaller than SN 1998bw 
(i.e., region A for $M_{\rm ms} = 25 M_{\odot}$). 
It is thus important to derive $M_{\rm ms}$ for these sub-luminous cases, 
by spectroscopic and light curve modeling, 
to distinguish these possibilities. 

On the other hand, the diversity in terms of $M_{\rm ms}$ is not important 
in interpreting no detection of supernovae associated with GRBs 060505 and 060614. 
If they are the outcome of the core collapse of 
a massive star, the properties of the wind/jet
should fall in the low end of $\dot E_{\rm w}$ and $\dot M_{\rm w}$ (region E), 
or region D with $\dot M_{{\rm iso}, \odot} \la 10^{-3}$. 
If this is not the case, we should have detected associated supernovae. 
Note that Tominaga et al. (2007b) showed that the wind/jet-driven explosion 
with small $\dot E_{\rm iso}$ "can" account for the non-detection of 
supernovae in these GRBs, while this study 
has shown that $\dot E_{\rm iso}$ and/or $\dot M_{\rm iso}$ "must" be small 
in the central engine of these GRBs. 
These GRBs highlight our suggestion that we can constrain the properties of 
the central engine by observations (non-detection, in this case) of 
associated supernovae. 

The present analysis is based on one-dimensional calculations. 
We predict that the different categories (A--E) are divided 
rather sharply in terms of $\dot M_{\rm iso}$ and $\dot E_{\rm iso}$, 
while in reality this may well be smoothed by the 
jet-accretion interaction (Maeda \& Nomoto 2003; Tominaga et al. 2007). 
Although the jet-accretion interaction should create a non-radial flow, 
key conclusions in the present paper are not defeated. 
For example, our model can explain the qualitative behaviors found 
in the past numerical study, and can reach at least rough quantitative agreement 
for the important quantity like $M$($^{56}$Ni) (for limited parameter space 
as investigated by numerical study; see also Maeda \& Nomoto 2003). 

The largest uncertainty involved in the present analysis is in the treatment 
of nucleosynthesis within the wind material, specifically, the assumption that 
20\% of the wind materials become $^{56}$Ni. Surman, McLaughlin, \& Hix (2006) concluded 
that the $^{56}$Ni mass fraction could be as large as $\sim 50$\% for the wind/jet 
with the entropy in the range between 10 and 30. This does not affect our conclusion 
for $\Gamma_{\rm w} \gg1$, as the mass of the wind materials is anyway 
small in this case. The only conclusion possibly affected by this uncertainty 
is the supernova feature in the region B. From equation 16, we see that the entropy 
of the wind materials falls into the range $s = 10 - 30$ if 
$\dot E_{{\rm iso},51} \dot M_{{\rm iso}, \odot} \sim 1$. 
This line crosses the vicinity of region B, and lead to the contribution of 
the wind materials for the $^{56}$Ni production comparable to that of  
the shocked mantle in this region. As such, $M$($^{56}$Ni) is this region 
may have a significant contribution from the wind materials.

\section*{Acknowledgments}
This research has been supported by World Premier 
International Research Center Initiative (WPI Initiative), 
MEXT, Japan, and partly by the Grant-in-Aid for Young Scientists 
of the JSPS/MEXT (20840007). 
N.T. is supported by the JSPS postdoctoral fellowship. 
The authors would like to thank Ken'ichi Nomoto, Hideyuki Umeda, 
Masaru Shibata, and Yuichiro Sekiguchi for useful discussion.

\bsp

\label{lastpage}

\end{document}